\documentclass[aip,amsmath,amssymb,reprint]{revtex4-2}
\usepackage[utf8]{inputenc}
\usepackage[T1]{fontenc}
\usepackage{mathptmx}
\usepackage{physics}
\usepackage{mathrsfs}
\usepackage{amsmath,gensymb}
\usepackage{amsfonts}
\usepackage{amssymb}
\usepackage{amsthm}
\usepackage{graphicx}
\usepackage{natbib}
\usepackage{color}
\usepackage{hyperref}
\usepackage{bm}
\usepackage{verbatim}
\usepackage{dcolumn,subfig,mathtools}
\usepackage{soul}
\usepackage[framemethod=tikz]{mdframed}

\begin{document}

\title{A Perspective on Magnon Spin Nernst Effect in Antiferromagnets}

\author{Hantao Zhang}
\affiliation{Department of Electrical and Computer Engineering, University of California, Riverside, CA 92521, USA}
\thanks{Correspondence should be sent to: hzhan289@ucr.edu}
\author{Ran Cheng}
\affiliation{Department of Electrical and Computer Engineering, University of California, Riverside, CA 92521, USA}
\affiliation{Department of Physics and Astronomy, University of California, Riverside, CA 92521, USA}

\begin{abstract}
Magnon excitations in antiferromagnetic materials and their physical implications enable novel device concepts not available in ferromagnets, emerging as a new area of active research. A unique characteristic of antiferromagnetic magnons is the coexistence of opposite spin polarization, which mimics the electron spin in a variety of transport phenomena. Among them, the most prominent spin-contrasting phenomenon is the magnon spin Nernst effect (SNE), which refers to the generation of transverse pure magnon spin current through a longitudinal temperature gradient. We introduce selected recent progress in the study of magnon SNE in collinear antiferromagnets with a focus on its underlying physical mechanism entailing profound topological features of the magnon band structures. By reviewing how the magnon SNE has inspired and enriched the exploration of topological magnons, we offer our perspectives on this emerging frontier that holds potential in future spintronic nano-technology.
\end{abstract}

\maketitle

Data processing, storage, and transmission with lower energy consumption without sacrificing the speed of operation has been a lasting call in the realm of electronics since the advent of very-large-scale integration circuits. Compared to conventional semiconductors using electrons to carry information, magnetic insulators in which information can be encoded by magnetic excitations emerge as a promising alternative because manipulating magnetic excitations does not involve physical motion of charges, hence obviating Joule heating at a fundamental level~\cite{chumak2015magnon,barman20212021}. At temperatures far below the magnetic ordering temperature, magnetic excitations manifest as propagating spin waves (dubbed magnons in the quantum picture) which carry quantized spin angular momenta against the equilibrium magnetization.

\begin{figure}
	\centering
  	\includegraphics[width=0.85\linewidth]{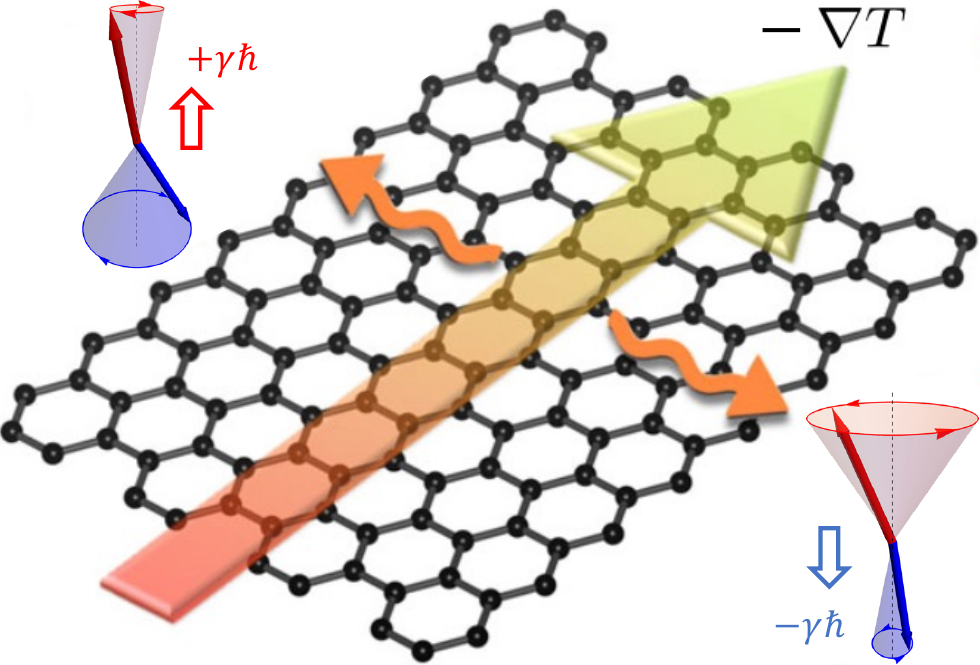}
	\caption{Schematic illustration of the magnon SNE and the two magnon species with opposite spin polarization. Here the gyro-magnetic ratio $\gamma>0$ so that the hollow arrows indicate the direction of non-equilibrium magnetic moments which are just opposite to their corresponding spin angular momenta.
	[Adapted from R. Cheng, S. Okamoto, and D. Xiao, Phys. Rev. Lett. $\bm{117}$, 217202 (2016).]}
	\label{fig:structure}
\end{figure}

In collinear antiferromagnetic (AFM) materials, however, magnetic order is characterized by the Néel vector whereas the macroscopic magnetization vanishes. Accordingly, there are two magnon species carrying opposite spin polarization, \textit{i.e.}, both up and down spins with respect to the Néel vector~\cite{keffer1953spin} as illustrated in Fig.~\ref{fig:structure}. The coexistence of both spin species in AFM magnons, forming an internal degree of freedom~\cite{cheng2016antiferromagnetic}, enables the magnonic realization of many physical phenomena associated with the electron spin~\cite{baltz2018antiferromagnetic,rezende2019introduction}. Since magnons are charge neutral and do not react directly to electric fields, thermal gradients usually act as the driving forces in magnonic spin-contrasting transport. For example, a temperature gradient can generate a pure magnon spin current longitudinally, known as the spin Seebeck effect~\cite{seki2015thermal,wu2016antiferromagnetic}. In the presence of the Dzyaloshinskii–Moriya (DM) interaction, which is an effective spin-orbit coupling for magnons, a temperature gradient can also produce a transverse magnon flow analogous to the anomalous Hall effect of electrons~\cite{Onose2010observation,Lee2010PRL}. Recent studies find that magnons in a two-dimensional (2D) ferromagnetic (FM) insulator with honeycomb lattice (such as monolayer CrI$_3$) are described by a bosonic version of Haldane's model thus exhibiting the magnon (thermal) Hall effect~\cite{owerre2016first,kim2016realization}. If such a honeycomb system is instead AFM (such as layered MnPS$_3$), the two spin species will move towards opposite transverse edges, leaving a pure spin current not accompanied by heat transport in the transverse direction. This phenomenon is known as the magnon spin Nernst effect(SNE)~\cite{cheng2016spin,zyuzin2016magnon} as illustrated in Fig.~\ref{fig:structure}, which realizes a magnonic analogy of the spin Hall effect and will be the central subject of review in this Perspective.

As will be detailed below, the magnon SNE in collinear AFM materials is far from a simple extension of the magnon Hall effect with the participation of opposite spin species. The profound topological structure of magnon dynamics revealed by the magnon SNE is distinct from its FM counterpart. While the prediction of magnon SNE aroused a series of theoretical generalizations under various physical contexts, heretofore it has only been measured experimentally in MnPS$_{3}$~\cite{shiomi2017experimental}. Nevertheless, the surging discovery of 2D magnets is rapidly enriching the material base for studying topological magnons, for which we anticipate a bright future in this intriguing direction of research.

In this Perspective, we first introduce the physical mechanism of the magnon SNE in an archetypal 2D AFM insulator (with honeycomb lattice, perpendicular Néel order, \textit{etc.}), followed by a discussion about the concomitant spin diffusion effect taking place in real devices which could substantially affect the electronic detection of the magnon SNE. Then we discuss the material candidates to realize the magnon SNE and the experimental progress along with practical challenges. We also discuss several recent developments pertaining to the physical origin of the magnon SNE in collinear and noncollinear AFM systems as well as paramagnets without long-range ordering.
%====================================

\begin{figure}
	\centering
  	\includegraphics[width=\linewidth]{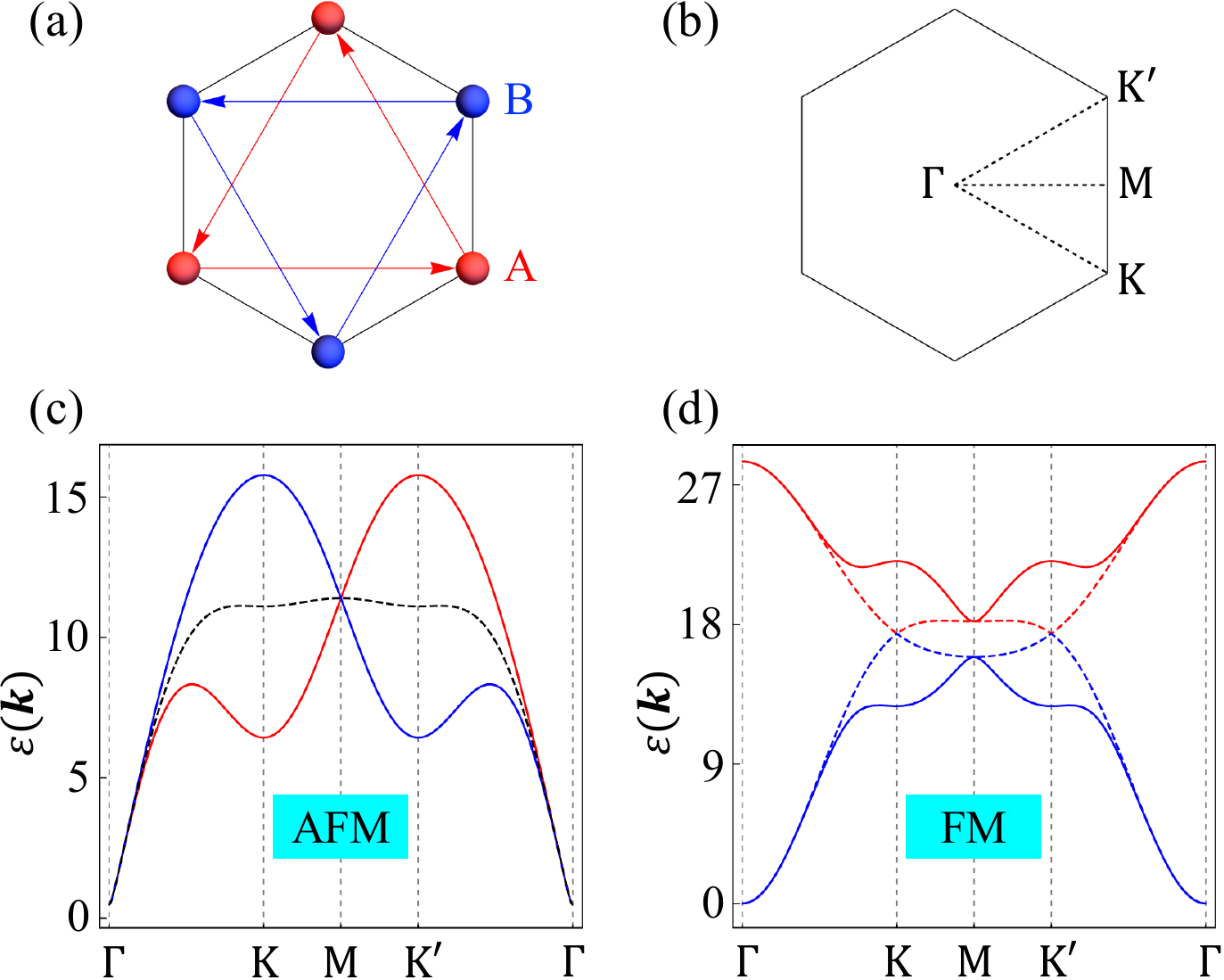}
	\caption{(a) Sublattice A and B in a hexagon. The sign of DM interaction takes $\xi_{ij}=1$ ($-1$) if $i,j$ in $\bm{S}_i\times\bm{S}_j$ is following (against) the arrows. (b) Brillouin zone and high symmetry points. The AFM (c) and FM (d) magnon bands solved from the same Hamiltonian Eq.~\eqref{eq:Hamiltonian} along the $\Gamma$-$K$-$K'$-$\Gamma$ loop for $D=0$ (dashed curves) and $D=0.1$ (solid curves). Parameters: $|J_{1}S| = 1.54$, $|J_{2}S| = 0.14$, $|J_{3}S| = 0.36$, and $K_{\rm eff}S = -0.0086$, all in meV.}
	\label{fig:AFM_FM_magnon}
\end{figure}

With layered MnPS$_3$ or its variances in mind, let us consider a 2D honeycomb AFM with collinear order normal to the plane, as illustrated in Fig.~\ref{fig:structure}. The spin dynamics of such a system can be effectively described by the Hamiltonian~\cite{cheng2016spin}
\begin{equation}\label{eq:Hamiltonian}
H = J_1 \sum_{\langle i,j\rangle} \bm{S}_{i} \cdot \bm{S}_{j} + D \sum_{\langle\langle i,j\rangle\rangle} \xi_{ij} \hat{\bm{z}} \cdot (\bm{S}_{i} \times \bm{S}_{j}) + K \sum_{i} S_{iz}^{2},
\end{equation}
where $\langle i,j\rangle$ and $\langle\langle i,j\rangle\rangle$ denote summations over the nearest and second-nearest neighboring bonds, respectively, $J_1 > 0$ is the exchange interaction, $K < 0$ is the easy axis anisotropy, $D$ is DM interaction along the $\hat{\bm{z}}$ direction (arising from the inversion symmetry breaking of second-nearest neighbors), and $\xi_{ij}=1(-1)$ when $\bm{S}_i$ and $\bm{S}_j$ are arranged in a counterclockwise (clockwise) manner as shown in Fig.~\ref{fig:AFM_FM_magnon}(a). The existence of easy axis anisotropy is essential for the stability of long-range N\'{e}el order by suppressing quantum fluctuations in ground state. To simplify our discussion, we ignore the exchange interactions among second and third neatest neighbors ($J_{2}$ and $J_{3}$) as they are one order of magnitude smaller than $J_1$ and do not affect the essential physics. However, we will reinstate these quantities in the numerical results later in order to reflect real material properties.

To find the magnon band structure, we invoke the linearized Holstein-Primakoff transformation~\cite{keffer1953spin,rezende2019introduction}
\begin{align}
 &S_{iA}^{+} = \sqrt{2S} a_{i},\quad S_{iA}^{-} = \sqrt{2S} a_{i}^{\dagger},\quad S_{iA}^{z} = S - a_{i}^{\dagger}a_{i}, \notag\\
 &S_{iB}^{+} = \sqrt{2S} b_{i}^{\dagger},\quad S_{iB}^{-} = \sqrt{2S} b_{i},\quad S_{iB}^{z} = -S + b_{i}^{\dagger}b_{i}, \notag
\end{align}
followed by Fourier transformations $\sqrt{N}a_i = \sum_{\bm{k}} \exp(i \bm{k} \cdot \bm{r}_{i})a_{\bm{k}}$ and $\sqrt{N}b_i = \sum_{\bm{k}} \exp(i \bm{k} \cdot \bm{r}_{i})b_{\bm{k}}$ with $S$ the spin magnitude and $N$ the total number of unit cells. Then Eq.~\eqref{eq:Hamiltonian} can be recast in a quadratic form as
\begin{align}\label{eq:Hk}
H &= \frac{S}{2} \sum_{\bm{k}} X_{\bm{k}}^\dagger
\begin{pmatrix}
A^{+}(\bm{k}) & 0 & 0 & J_1 f(\bm{k}) \\
0 & A^{+}(\bm{k}) & J_1 f^{*}(\bm{k}) & 0 \\
0 & J_1 f(\bm{k}) & A^{-}(\bm{k}) & 0 \\
J_1 f^{*}(\bm{k}) & 0 & 0 & A^{-}(\bm{k})
\end{pmatrix} 
X_{\bm{k}},
\end{align}
where $X_{\bm{k}} = [a_{\bm{k}}, b_{\bm{k}}, a^{\dagger}_{-\bm{k}}, b^{\dagger}_{-\bm{k}}]^{T}$ is the Nambu basis~\footnote{An equivalent choice of the Nambu basis is $X_{\bm{k}}=[a_{\bm{k}}, b_{\bm{k}}, a^{\dagger}_{\bm{k}}, b^{\dagger}_{\bm{k}}]^{T}$, which requires the Fourier transformation on the $B$ sublattice to be $\sqrt{N}b_i = \sum_{\bm{k}} \exp(-i \bm{k} \cdot \bm{r}_{i})b_{\bm{k}}$}, $A^{\pm}(\bm{k}) = 3J_{1} - K (2S - 1)/S \pm D g(\bm{k})$, $f(\bm{k}) = \sum_{i} \exp{(i \bm{k} \cdot \bm{d}_{i})}$ and $g(\bm{k}) = 2\sum_{i \in \text{odd}} \sin{(\bm{k} \cdot \bm{a}_{i})}$ with $\bm{d}_i$ and $\bm{a}_i$ being the vectors of the first and second nearest neighboring links. By performing a Bogoliubov transformation, one can diagonalize the Hamiltonian and obtain four different bands, where two of them are redundant solutions. The two physical solutions are
\begin{equation}\label{eq:AFM_magnon_spectrum}
\epsilon_{\pm}(\bm{k})/S = \sqrt{(3J_{1} -K_{\rm eff})^{2} - |J_{1} f(\bm{k})|^{2}} \pm Dg(\bm{k}),
\end{equation}
where $K_{\rm eff}=K(2S-1)/S$ is the effective anisotropy and $\pm$ distinguishes the two spin species. When retrieving the ignored $J_2$ and $J_3$, the first term on the right-hand side of Eq.~\eqref{eq:AFM_magnon_spectrum} becomes very complicated, but it always satisfies $c_6$ ($6$-fold rotational) symmetry in the Brillouin zone. It will become clear later that the magnon SNE is enabled by the $\pm Dg(\bm{k})$ term which respects $c_3$ symmetry in the Brillouin zone and is not affected by $J_2$ and $J_3$. Using material parameters of MnPS$_3$, we plot the magnon bands numerically along the path of $\Gamma$-$K$-$M$-$K'$-$\Gamma$ [depicted in Fig.~\ref{fig:AFM_FM_magnon}(b)] in Fig.~\ref{fig:AFM_FM_magnon}(c), where the DM interaction lifts the degeneracy of bands in opposite fashions. It should be noted that the magnon bands can also be solved by linearizing the Landau-Lifshitz equation without resorting to the quantum formalism.

As a comparison, we plot in Fig.~\ref{fig:AFM_FM_magnon}(d) the magnon bands of a 2D honeycomb FM insulator described by exactly the same Hamiltonian with only $J_1$ flipped sign, which, when $J_2$ and $J_3$ are ignored, reduces to
\begin{equation}\label{eq:FM_magnon_spectrum}
\epsilon(\bm{k})/S = -3J_1-K_{\rm eff} \pm \sqrt{|J_{1} f(\bm{k})|^{2} + |Dg(\bm{k})|^{2}}.
\end{equation}
Different from the AFM case, the two FM magnon bands have the same spin polarization. If the DM interaction vanishes, the two bands are degenerate at the $K$ and $K'$ points where they form two Dirac cones analogous to the Dirac electrons in graphene. The DM interaction opens gaps at the Dirac points, leading to a topological phase transition from a trivial magnon insulator to a topological magnon insulator~\cite{owerre2016first,kim2016realization,wang2021topological}. In the AFM case, by contrast, the magnon bands of opposite spin polarization are deformed in just opposite ways around the $K$ and $K'$ valleys because of the DM interaction, leaving the bands degenerate only at the $\Gamma$ and $M$ points. Since the total spin $S^{z} = \sum_{i} S^{z}_{i}=\sum_{i}(b_k^\dagger b_k-a_k^\dagger a_k)$ commutes with the Hamiltonian, $\langle S^{z}\rangle=\pm\hbar$ is conserved for each band. This situation is quite different from the spin Hall effect of electrons, where none of the spin component is conserved.

\begin{figure}
	\centering
  	\includegraphics[width=\linewidth]{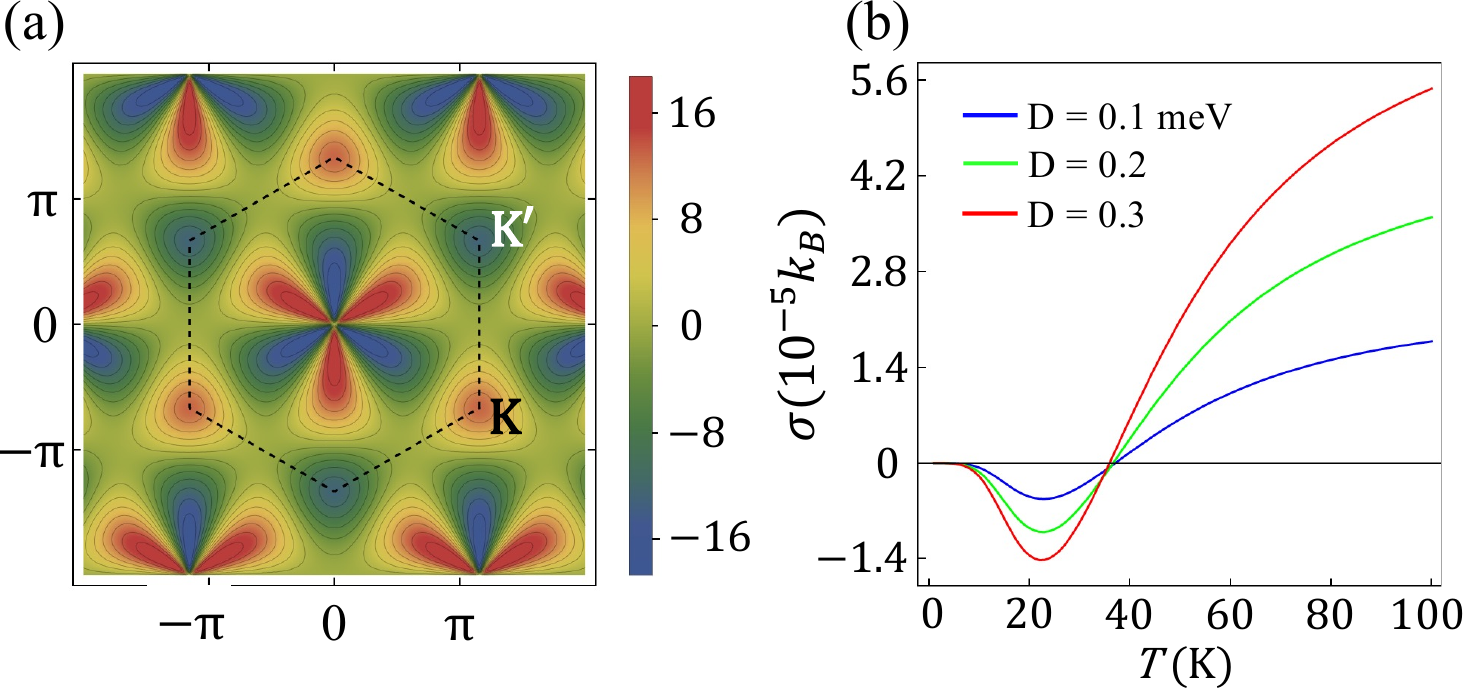}
	\caption{(a) Berry curvature of the spin-down (right-handed) AFM magnons for $D = 0.36$meV. Dashed lines enclose the first Brillouin zone. (b) Spin Nernst coefficient as a function of temperature for different strengths of the DM interaction. The values of $|J_{1}|$, $|J_{2}|$, $|J_{3}|$, $K_{eff}$ follow those in Fig.~\ref{fig:AFM_FM_magnon}. [Adapted from R. Cheng, S. Okamoto, and D. Xiao, Phys. Rev. Lett. $\bm{117}$, 217202 (2016).]}
	\label{fig:eng_Berry}
\end{figure}

What drives the transverse motion of magnons is the Berry curvature acting as an artificial magnetic field in the momentum space. In our context, the magnon Berry curvature is defined as
\begin{equation} \label{eq:Berry_curvature}
\bm{\Omega}_{\pm}(\bm{k}) = i \bm{\nabla}_{\bm{k}} \times  \left[ 
\bra{\psi_\pm(\bm{k})} (\sigma_{z} \otimes \tau_0)  \bm{\nabla}_{\bm{k}} \ket{\psi_\pm(\bm{k})} \right], 
\end{equation}
where $\pm$ corresponds to the band $\varepsilon_{\pm}$ with wavefunction $\ket{\psi_\pm(\bm{k})}$, $\nabla_{\bm{k}}$ is the gradient over $\bm{k}$, $\sigma_{z}$ is the Pauli matrix in the spin space, $\tau_0$ is the identity matrix in the pseudospin space ($A-B$ sublattices), and $\otimes$ is the tensor product. In 2D systems, $\bm{\Omega}_\pm(\bm{k})$ reduces to a pseudo scalar that only has a $z$ component. While at any $\bm{k}$ point the Berry curvatures from the four bands directly solved from the Hamiltonian add up to zero, in the subspace of physical solutions, however, we have $\Omega_+(\bm{k})=\Omega_-(\bm{k})$, \textit{i.e.}, the spin-up and spin-down bands share the same Berry curvature, which is plotted in Figure~\ref{fig:eng_Berry}(a). We see that the Berry curvature is an odd function of $\bm{k}$ and, in the vicinity of $K$ and $K'$ points, it shows opposite values. Driven by an in-plane temperature gradient $\nabla T$, the transverse magnon current contributed by each magnon band reads~\cite{Matsumoto2011PRL,cheng2016spin}
\begin{align} \label{eq:magnon_current}
\bm{j}_\pm &= \frac{k_{B}}{\hbar} \hat{\bm{z}} \times \bm{\nabla}T \int \frac{d^{2}k}{(2\pi)^{2}} \bm{\Omega}_\pm(\bm{k}) \{n_{\pm}(\bm{k}) \ln n_{\pm}(\bm{k}) \notag \\ 
&\qquad\qquad\qquad - [1 + n_{\pm}(\bm{k})] \ln [1 + n_{\pm}(\bm{k})] \}, 
\end{align}
where $n_{\pm}(\bm{k}) = 1/ \{\exp[\epsilon_{\pm}(\bm{k}) / k_{B} T] - 1\}$ is the Bose-Einstein distribution function with $k_{B}$ the Boltzmann constant.

Given the total spin current: $\bm{j}_s=\hbar(\bm{j}_+-\bm{j}_-)\equiv-\sigma\hat{\bm{z}} \times \bm{\nabla}T$, the SNE coefficient can be defined as $\sigma=\sigma_+-\sigma_-$, which is plotted in Fig.~\ref{fig:eng_Berry}(b) as a function of temperature for three different values of the DM interaction. In sharp contrast to the FM case in which non-zero Berry curvature arises only when the DM interaction is non-zero, the Berry curvature in the AFM case, quite counter-intuitively, is \textit{independent} of the DM interaction; it is a quantity intrinsic to the Néel ordering on a honeycomb lattice. The DM interaction imbalances the magnon population between the $K$ and $K'$ points (the "valleys" in the Brillouin zone) by changing their energy distribution according to their spin polarization. Specifically, the spin-up (spin-down) band develops a local minimum at the $K$ ($K'$) point and a local maximum at the $K'$ ($K$) point as illustrated by the red (blue) curve in Fig.~\ref{fig:AFM_FM_magnon}(c). Because of the Bose-Einstein distribution, the spin-up (spin-down) magnons are preferably affected by positive (negative) Berry curvature shaded in red (blue) color in Fig.~\ref{fig:eng_Berry}(a), hence deflecting towards a preferred transverse edge. As a comparison, we notice that in the FM case, the Berry curvature of the lower band does not change sign in the Brillouin zone and is the same at both $K$ and $K'$. In the upper band, the Berry curvature is of equal magnitude and opposite sign at all $\bm{k}$ points. Therefore, magnons from the lower and upper bands move in opposite transverse directions with an imbalanced thermal population, resulting in a net thermal Hall current.

There exists a different convention in reducing the four bands to the physical solution, which ends up with two degenerate bands $\epsilon(\bm{k})/S = \sqrt{(3J_{1} -K_{\rm eff})^{2} - |J_{1} f(\bm{k})|^{2}} + Dg(\bm{k})$ for both spin species even when $D\neq0$. In that case, however, magnons of opposite spin polarization are subject to opposite Berry curvatures, \textit{i.e.}, $\bm{\Omega}_+(\bm{k})=-\bm{\Omega}_-(\bm{k})$. Therefore, Eq.~\eqref{eq:magnon_current} still gives $\bm{j}_+=-\bm{j}_-$, which is exactly the same as what we obtained in our convention. While the magnon SNE originates from the Berry curvature, the topological Chern number of each magnon band is exactly zero because $\Omega_{\pm}(\bm{k})=-\Omega_{\pm}(-\bm{k})$, and there is no topologically protected edge state in a monolayer sample. However, a bilayer honeycomb AFM insulator in which the degree of freedom get doubled can afford topologically protected edge magnons~\cite{zyuzin2016magnon}. Similar to the topological insulator, one can further define the $\mathbb{Z}_{2}$ number to characterize the magnon SNE, which coincides with the spin Chern number as $S_z$ is conserved~\cite{kondo2019z}.

As shown in Fig.~\ref{fig:eng_Berry}(b), a striking feature of the SNE coefficient is the non-monotonic temperature dependence and the sign change, which is also indicated by a recent experiment~\cite{shiomi2017experimental}. This exotic behavior can be attributed to the competition between magnons near the $\Gamma$ point and those from the $K$ and $K'$ valleys, because the Berry curvature near these regions has opposite values as shown in Fig.~\ref{fig:eng_Berry}(a). Specifically, the $\pm Dg(\bm{k})$ term in Eq.~\eqref{eq:AFM_magnon_spectrum}, which is an odd function of $\bm{k}$, is responsible for the spin-dependent imbalance of the magnon population, hence the ensuing magnon SNE. As shown in Fig.~\ref{fig:AFM_FM_magnon}(c), this term culminates at the $K$ and $K'$ points while it approaches zero near the $\Gamma$ point. On the other hand, the Berry curvature is more prominent around the $\Gamma$ point as compared to the valleys. Consequently, at very low temperatures where valley magnons are not sufficiently populated due to their relatively high energy, $\Gamma$-magnons dominate the SNE even though the magnon bands are almost rotationally symmetric near $\Gamma$. With an increasing temperature, however, the valley magnons become more and more important, which will eventually overwhelm the $\Gamma$-magnons thanks to the significant band asymmetry around $K$ and $K'$. We point out that although the competition between $\Gamma$-magnons and valley magnons is universal, whether a sign reversal of the SNE coefficient can be observed is very material specific and may not be guaranteed. 

%\begin{mdframed}[hidealllines=true,backgroundcolor=yellow]
%\end{mdframed}

Similar to the SNE of electrons~\cite{meyer2017observation,sheng2017spin}, we can define the SNE angle $\theta_{SN}$ as the ratio of the magnon SNE current over the longitudinal magnon current as $\theta_{SN} = (j_{+} - j_{-}) / j_{x}$, where $j_x$ is the longitudinal magnon current (assuming $\bm{\nabla}T$ is applied along the $x$ direction)~\cite{rezende2019introduction}:
\begin{align} \label{eq:SS_magnon_current}
j_{x} = \frac{\partial T}{\partial x} \int \frac{d^{2}\bm{k}}{(2\pi)^{2}} \left[ \tau_{+} v_{+}^{2} \frac{\partial n_{+}(\bm{k})}{\partial T}+\tau_{-}v_{-}^{2} \frac{\partial n_{-}(\bm{k})}{\partial T} \right], 
\end{align}
where $\tau_{\pm}$ is the phenomenological magnon relaxation time and $v_{\pm} = \partial \epsilon_{\pm}(\bm{k}) /\hbar \partial k_{x}$ is the magnon group velocity. While $\theta_{SN}$ for magnons is conceptually parallel to its electronic counterpart, we are not able to estimate its value in MnPS$_3$ and other candidate materials, because converting the magnon Seebeck current (which is a heat current) into $j_x$ is rather difficult, let alone that the magnon Seebeck effect is itself an open question in 2D magnets.

Similar to the spin Hall effect of electrons, here the absence of thermal Hall effect ($\sigma_+=-\sigma_-$, so $\bm{j}_++\bm{j}_-=0$) is protected by the time reversal symmetry. A finite thermal Hall current can be induced by applying a perpendicular magnetic field $B$. However, the $B$ field only shifts the dispersion relations by an overall constant depending on the spin polarization while leaving the eigenstates and the Berry curvature unchanged. In the presence of $B$ field, $\sigma_+(B) = \sigma/2 + \Delta \sigma_{1}$ and $\sigma_-(B) = -\sigma/2 + \Delta \sigma_{2}$ where $\Delta \sigma_{1}$ is different from $\Delta \sigma_{2}$. Accordingly, the SNE coefficient becomes $\sigma(B)=\sigma_+(B)-\sigma_-(B)=\sigma +\Delta\sigma_1-\Delta\sigma_2$. If $B$ is reversed, symmetry dictates that $\sigma_+(-B) = \sigma/2 - \Delta \sigma_{2}$ and $\sigma_-(-B) = -\sigma/2 + \Delta \sigma_{1}$, which yields $\sigma(-B)=\sigma(B)$, \textit{i.e.}, the SNE coefficient is an even function of $B$. Consequently, to the lowest order, the transverse spin current $\bm{j}_{s}$ must scale as $B^{2}$. In other words, while a perpendicular $B$ field can induce a net magnetization on top of the Néel order by polarizing the equilibrium magnons, it only introduces a higher order effect in the non-equilibrium magnon transport.

\begin{figure}
	\centering
  	\includegraphics[width=\linewidth]{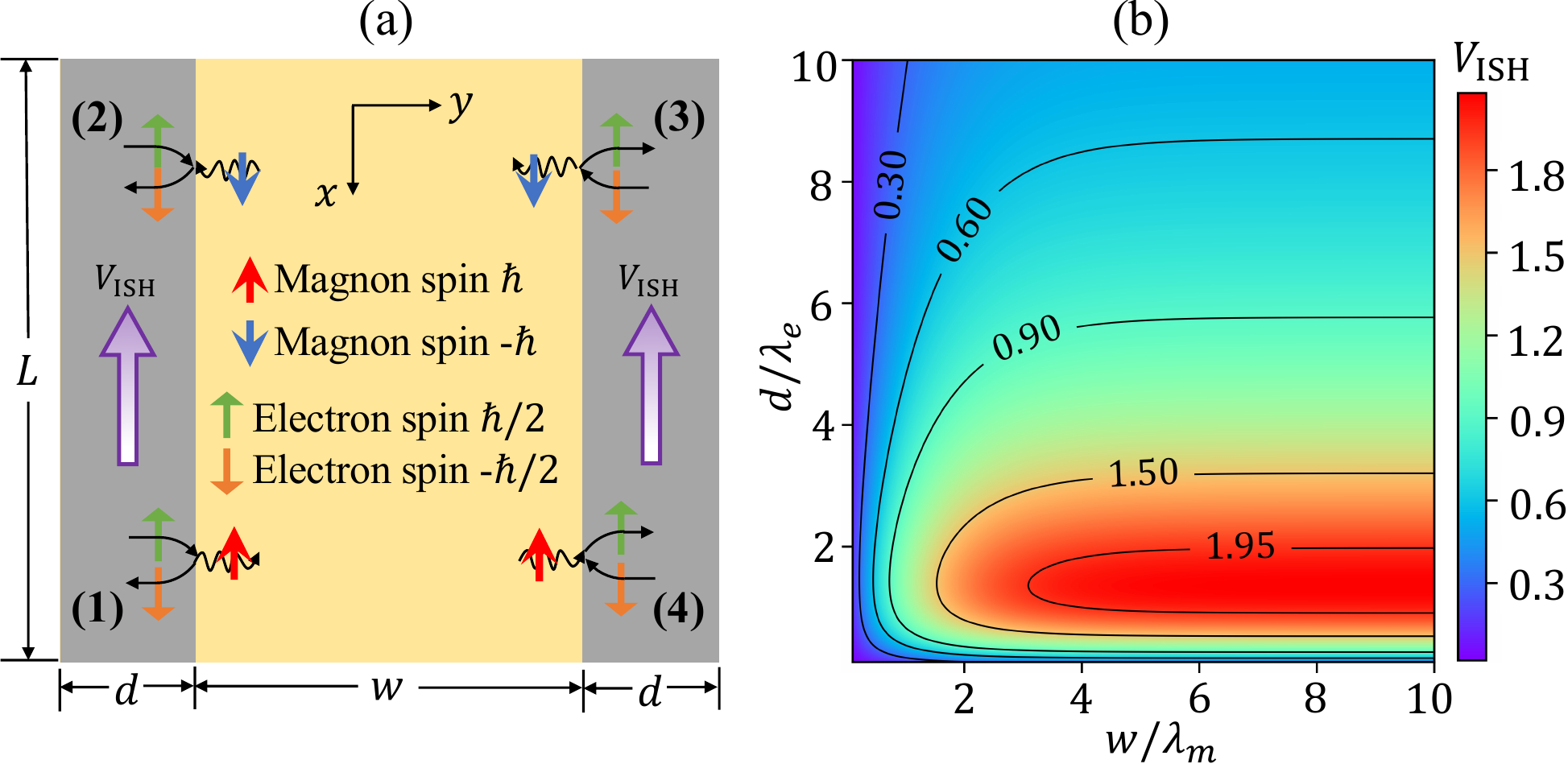}
	\caption{(a) Schematics of a lateral device measuring the magnon SNE, where a temperature gradient along $\hat{\bm{x}}$ drives a magnon spin current flowing along $\hat{\bm{y}}$. Spin conversions between magnons and electrons take place on the boundaries through four different scattering processes sketched as (1) -- (4). Spin currents injected into the leads eventually convert into charge voltage along $\hat{\bm{x}}$ through the inverse spin Hall effect. (b) Inverse spin Hall voltage in the unit of $-\partial_{x}T(\sigma \theta_{s} L/eG\lambda_{e})$ as a function of the AFM insulator width $w/\lambda_{m}$ and the lead width $d/\lambda_{e}$. The diffusion factors are taken to be$\eta_{m} = \eta_{e} = 16$. [Adapted from  H. Zhang and R. Cheng, Phys. Rev. Appl. $\bm{16}$, 034035 (2021).]}
	\label{fig:diffusion_model_phase}
\end{figure}

The magnon SNE was originally proposed as a ballistic effect without considering the influence of boundaries and spin diffusion~\cite{cheng2016spin,zyuzin2016magnon}. However, in real experiments~\cite{shiomi2017experimental}, boundaries and spin diffusion can be essential. A diffusive theory of the magnon SNE is formulated in Ref.~\cite{zhang2021spin}, here we only demonstrate the key ideas without demonstrating the technical details. Figure~\ref{fig:diffusion_model_phase}(a) displays a typical experimental setup for the electronic detection of the magnon SNE, in which an AFM layer is in contact with two heavy metallic leads on both sides. A temperature gradient (along $x$ direction) generates a pure magnon spin current (along $y$ direction) that diffuses towards the leads, injecting spin angular momenta across the boundaries which eventually are converted into a charge voltage $V_{ISH}$ through the inverse spin Hall effect. The spin injection processes are illustrated by Feynman diagrams labeled (1)--(4) in Fig.~\ref{fig:diffusion_model_phase}(a), each of which involves a spin-flip scattering of an electron and the creation or annihilation of a magnon, satisfying the spin conservation.

Solving the spin diffusion equations on both sides of each boundary with boundary conditions reflecting the four spin-dependent scattering processes depicted in Fig.~\ref{fig:diffusion_model_phase}(a) yields an inverse spin Hall voltage as
\begin{eqnarray} \label{eq:ISH_V_no_field}
&&V_{ISH} = -\partial_x T\frac{\sigma\theta_{s} L}{2eGd} \frac{\eta_m\eta_e\tanh\frac{d}{2\lambda_e}}{\eta_m+\coth\frac{w}{2\lambda_m}\left(1+2\eta_e\coth\frac{d}{\lambda_e}\right)}, 
\end{eqnarray}
where $\partial_{x} T$ is the temperature gradient along $x$ direction, $\theta_{s}$ is the spin Hall angle, $e$ is the electron charge, $L$, $w$ and $d$ describe the device geometry shown in Fig.~\ref{fig:diffusion_model_phase}(a), $\eta_{m} = G\lambda_{m} / D_{m}$ and $\eta_{e} = G\lambda_{e} / D_{e}$ with $G$ being the phenomenological interfacial spin conductance and $\lambda_{m}$ ($\lambda_{e}$) and $D_{m}$ ($D_{e}$) being the spin diffusion length and spin diffusivity for magnons (electrons), respectively. Figure~\ref{fig:diffusion_model_phase}(b) plots Eq.~\eqref{eq:ISH_V_no_field} versus the AFM and the lead widths using material parameters of MnPS$_{3}$, where we identify a dramatic amplification of the detectable signal $V_{ISH}$ (orders of magnitude larger) when $d\sim\lambda_e$ and $w\gg\lambda_m$ as compared to the ballistic limit $w\ll\lambda_m$. In other words, the magnon diffusion can remarkably facilitate the electronic detection of the magnon SNE. While the diffusion effect magnifies the output signal, it does not change the system symmetry such that $V_{ISH}$ still scales as $B^{2}$, which makes it challenging to separate the magnon SNE from other effects, especially the thermoelectric voltage. Non-local transport measurement based on Hall bar devices can potentially overcome this problem, because the magnon spin generation is spatially isolated from the spin detection. A similar idea is to insert heat-insulating but spin-conducting materials between the leads and the AFM region in Fig.~\ref{fig:diffusion_model_phase}(a). Aside from non-local experiments, one can resort to the inverse SNE of magnons to identify the prediction, namely, detecting the transverse temperature drop induced by a longitudinal pure spin injection.

Concerning reliability, we believe that optical detection, which is devoid of thermoelectric effects, should be superior to electronic detection so long as the spatial resolution is sufficient to resolve the non-equilibrium magnon distribution in the transverse direction. Without metallic leads, the magnon SNE will generate opposite spin accumulations on opposite transverse edges, and the detection scheme should be similar to the optical detection of the spin Hall effect~\cite{kato2004observation}. We notice that the recently developed nitrogen-vacancy center magnetometer~\cite{du2017control} is suitable for this purpose thanks to its high sensitivity and spatial resolution.

Until now, we are aware of only one experimental study of the magnon SNE~\cite{shiomi2017experimental}, where a thin-film MnPS$_{3}$ is used and the experimental setup is similar to that in Fig.~\ref{fig:diffusion_model_phase}(a). While they indeed observed a non-monotonic temperature dependence of $V_{ISH}$, it is not decisive whether the signal truly stems from the magnon SNE because the thermoelectric voltage could not be unambiguously separated from the overall $V_{ISH}$ and the dependence on a perpendicular magnetic field was not measured. A more carefully designed measurement is needed to identify the predicted magnon SNE, and as mentioned above, optical detection is more preferable to electronic detection.

Following the investigation of MnPS$_{3}$ as a promising candidate to verify the magnon SNE~\cite{cheng2016spin}, there has been several transition metal trichalcogenides (TMT) predicted to exhibit similar effects. A recent DFT study found that MnPSe$_{3}$ and VPS$_{3}$, both with N\'eel AFM ground state, could show a stronger magnon SNE compared to that of MnPS$_{3}$~\cite{bazazzadeh2021symmetry}. Moreover, we notice that FePS$_{3}$~\cite{lee2016ising}---a close cousin of MnPS$_{3}$ in the TMT family, should be another good candidate. Unlike MnPS$_{3}$, FePS$_{3}$ has been successfully made into monolayer, in which the theoretical modeling is more applicable. Besides the Néel order, a TMT magnet can host very rich collinear magnetic orders such as stripe, zigzag and armchair states depending on the relations among $J_1$, $J_2$ and $J_3$~\cite{Sivadas2015PRB,boyko2018evolution}. According to recent studies, the magnon SNE is not exclusive to the Néel ground state. For example, both the stripe phase and the zigzag phase are found to exhibit the magnon SNE~\cite{lee2018magnonic}. On the materials side, zigzag AFM insulators such as CrSiTe$_{3}$, NiPS$_{3}$ and NiPSe$_{3}$ are predicted to afford even larger SNE coefficients compared to that in MnPS$_{3}$~\cite{bazazzadeh2021symmetry}. In Ref.~\cite{lee2018magnonic}, the SNE coefficient of the zigzag phase undergoes a sign change with rising temperature, but none of the zigzag AFM materials show this pattern in Ref.~\cite{bazazzadeh2021symmetry}. This discrepancy may be attributed to the different parameters used in different studies, which may significantly affect the temperature dependence as we have commented in the discussion of Fig.~\eqref{fig:eng_Berry}(b).

Moreover, these non-Néel phases can have Dirac-like nodes~\cite{boyko2018evolution}, which are otherwise prohibited in the N\'eel phase. Going beyond 2D honeycomb lattice, Ref.~\cite{zyuzin2018spin} extends the study of Weyl magnons and the associated SNE to a structure composed of stacked FM and AFM layers, where the DM interaction splits the Dirac node into Weyl points. If the 2D honeycomb layers are stacked and coupled antiferromagnetically, magnons in the N\'eel state will have Dirac nodal lines robust against the DM interaction. In this case, however, the SNE vanishes because the contribution from each layer cancels out~\cite{owerre2019topological}. In addition to the honeycomb lattice, a 2D checkboard lattice is shown to host the magnon SNE as well, where the Chern number of each magnon band is zero even in the presence of DM interaction~\cite{pires2019magnon,lima2021spin}. Furthermore, the magnon SNE in 3D magnets (non-Van del Waals materials) remains a fully uncharted territory. Given that the magnon Hall effect was first discovered in 3D rather than 2D
~\cite{Onose2010observation}, it is tempting to ask if the magnon SNE can take place in 3D materials, which may exhibit complicated non-N\'{e}el spin structures.

In the above, the magnon SNE and its manifestations in various AFM materials all requires the system to be long-range ordered. However, recent studies found that the SNE phenomenon also manifests in the disordered paramagnetic phase above the N\'eel temperature~\cite{kim2016realization,zhang2018spin}. At first glance, this is counter-intuitive because in the paramagnetic phase magnons are not even defined; spin fluctuations from an otherwise non-magnetized ground state play the role of spin carriers. In contrast to the ordered phase, the paramagnetic SNE exists in both FM and AFM materials, where the thermal Hall current vanishes in the absence of magnetic fields. This fact indicates that the honeycomb lattice structure along with the symmetry of magnetic interactions, especially the second-nearest neighboring DM interaction, is quintessential to the SNE phenomenon whereas the long-range magnetic order is not the most fundamental requirement as one would otherwise expect. The temperature dependence of paramagnetic SNE coefficient is non-monotonic in both the AFM and FM cases, thanks to the increasing population of spin fluctuations and the decreasing mean-field order parameters with increasing temperature. So far, whether the paramagnetic SNE exists in other types of lattice is unclear, awaiting further investigation.

Another interesting playground of the magnon SNE is AFM insulators with noncollinear ground states. For example, KFe$_3$(OH)$_6$(SO$_4$)$_2$ has a Kagome lattice with in-plane $120\degree$ spin alignment. A single layer sample can exhibit the magnon SNE~\cite{mook2019spin,li2020intrinsic}. However, different from the collinear case, the magnon spin polarization in a noncollinear Kagome magnet can be either out-of-plane or in-plane, leading to two kinds of magnon SNE. In Ref.~\cite{li2020intrinsic}, the out-of-plane polarized SNE is weaker than in-plane polarized SNE, while in Ref.~\cite{mook2019spin} the out-of-plane component is not visible in simulation. While the magnon SNE in noncollinear magnets still originates from the Berry curvature, the underlying topological structure is more complicated than that of the collinear magnets~\cite{Bonbien2021}. There are three magnon bands in a noncollinear Kagome magnet consisting of three distinct sublattices, where the band Chern numbers are -3, 1 and 2~\cite{li2020intrinsic}. Furthermore, the DM interaction is believed to induce an out-of-plane canting of spins, which gives rise to a small net magnetization in the ground state. But it turns out that even without the DM interaction, the SNE can still exist~\cite{mook2019spin}, in stark contrasts with the collinear case. So far, a diffusive description of the magnon SNE in noncollinear magnets is absent.

Finally, we would like to point out that physical mechanisms other than the DM interaction could be responsible for the magnon SNE even in the simple collinear case. References~\cite{park2020thermal,bazazzadeh2021magnetoelastic,zhang20203} showed that by virtue of magnetoelastic coupling, phonons can hybridize with magnons and a non-zero Berry curvature emerges near the anti-crossing regions of the phonon and magnon bands. This phenomenon should be general and the induced Berry curvature leads to both the magnon Hall effect and the magnon SNE. In particular, the magnon-phonon hybridization can lead to topological magnetoelastic bands of higher Chern nuber~\cite{ma2021dzyaloshinskii}. However, the form of magnetoelastic coupling is different in these studies, and only square and honeycomb lattices are discussed, leaving wide room for theoretical extensions. As for the importance of branching out for alternative mechanisms to interpret the magnon SNE, we notice that a recent neutron scattering measurement of MnPS$_{3}$~\cite{wildes2021search} found no visible difference in the magnon energy between the $K$ and $K'$ points, suggesting a negligible DM interaction contrary to what we believed previously. Nonetheless, other TMTs in this family of materials, especially those with heavy elements, are still potential candidates for the magnon SNE. We also notice that a recent experiment confirmed the existence of Kitaev's interaction in 2D honeycomb FM CrI$_3$~\cite{Lee2020Kitaev}, which can lead to gap opening at $K$ and $K'$ points shown in Fig.~\ref{fig:AFM_FM_magnon}(d) even without the DM interaction. As the symmetry condition for Kitaev's interaction is very general, it is natural to ask if honeycomb AFM materials also afford non-negligible Kitaev's interaction, and, whether this interaction can lead to the magnon SNE along with other exotic spin-contrasting transport effects.

Similar to the magnon-phonon hybridization, other forms of quasi-particle hybridization may also enable the magnon SNE so long as the hybridized bands with anti-crossing gaps are topologically different from the non-hybridized bands. For example, the magnon SNE has been recently extended into the nonlinear response regime in which the spin Hall current of magnons is proportional to $(\nabla T)^2$~\cite{kondo2021nonlinear}. Moreover, the dipole-dipole interaction can enable the magnon SNE~\cite{shen2020magnon} in a 2D AFM system without the DM interaction. While the DM interaction is local, the dipolar interaction is non-local and long-ranged, which becomes increasingly important for larger systems.

In summary, we discussed the underlying mechanism and the topological nature of the magnon SNE in 2D collinear AFM insulators and how its detection is affected by spin diffusion. The pioneering studies of the magnon SNE in MnPS$_{3}$ inspired fruitful investigations into various materials and more complicated structures, including other mechanisms than the DM interaction. In spite of plenty theoretical extensions, the magnon SNE has so far been measured experimentally in very limited materials, calling for more experimental studies in the near future which not only provides more insights into the physical origin but also facilitates the development of practical magnonic devices.

This work is supported by the Air Force Office of Scientific Research under grant FA9550-19-1-0307. The authors are grateful to A. Balandin, D. Xiao, S. Okamoto and X. Chen for helpful discussions.

\bibliography{manuscript_of_Diffusive_spin_Nernst_effect}

\end{document}